# Second-harmonic generation due to coulomb-like interaction in a heterodimer of subwavelength dimensions

*Maya H. Shor[1, ‡], Esti Toledo[2, ‡], Shilpi Shital[1], Achyut Maity[1], Yonatan Sivan[3], Mark Schvartzman[2], and Avi Niv[1*]*

1. Department of Solar Energy and Environmental Physics, Jacob Blaustein Institutes for Desert Research, Ben-Gurion University of the Negev, Israel.
2. Department of Materials Engineering, Ben-Gurion University, Israel.
3. School of Electrical and Computer Engineering, Ben-Gurion University of the Negev, Israel.

**Abstract**

We experimentally study the optical second harmonic generation (SHG) from deep subwavelength gold-silver heterodimers and silver-silver and gold-gold homodimers. Our results show that the SHG from the heterodimer is about an order of magnitude more intense. In contrast, calculations based on known theory suggest that it is the silver-silver homodimer that should have the upper hand, putting the present theoretical view at odds with our experimental findings. Following this observation, we propose a dynamic model where nonlinearity emerges not from the material particles themselves but from the Coulombic interaction between them that prevails over other nonlinear optics mechanisms at the extreme subwavelength dimensions of the dimer in this case. The model's good agreement, combined with the failure of the conventional theoretical view, implies a yet uncharted nonlinear optical effect.

**Introduction**

Nonlinear optics of nanoscaled systems is a well-established field where second-order processes in isolated particles or dimers,[1–4] as well as other arrangements,[5–7] have all been studied. Yet, little attention was given to heterodimers - a dimer system of different material particles. Having a well-known linear response,[8–12] heterodimers are also appealing for second-order optical nonlinearity due to the asymmetry they may possess.[13] Here, we investigate the second harmonic generation (SHG) from extreme nanoscaled silver-gold (Au-Ag) heterodimer and homodimers (Ag-Ag and Au-Au) that were fabricated with accurately controlled nanometric scale inter-particle spacing using state-of-the-art nanoimprint lithography and double-angle metal deposition.

SHG is a process where two photons of the excitation light produce a single photon at twice the original frequency. While being forbidden within the metal volume due to the centrosymmetric nature of the crystal, it is allowed at their surface due to the following inversion symmetry violating mechanisms: Discontinuity in the screening potential spanning few angstroms from the interface and the finite electromagnetic penetration depth spanning few tens of nanometers into the metal at optical frequencies - Two mechanisms that are respectively known as the *surface* and *bulk* sources of SHG from a flat metal surface.[14–17] Over the years, these same mechanisms were invoked to explain the SHG from small metal-spheres and other complex-shaped nanostructures, with the only difference of being applied to the curved surface of the nanoparticle.[18–22] In the context of this article, the above is the *conventional-view* of such processes.

In this work, we have experimentally studied SHG from subwavelength nanoscaled gold (Au) and silver (Ag) homo- and hetero-dimers. The heterodimers displayed a marked spectral peak that was not present in the homodimers' response. Moreover, the peak SHG of the heterodimers was about an order of magnitude more intense than either the Ag or Au homodimers. This outcome was surprising given that the simulation of the conventional-view of such processes came to good agreement with the homodimer case only. Moreover, according to the simulation, it was the Ag-Ag homodimer that should have had the upper hand in terms of the SHG intensity for a given source power, which was not the case. Finally, as far as homodimers were concerned, the simulations were in good agreement with previous findings.[18,20–22] We have concluded, therefore, that the conventional-view of nonlinear metal optics falls short of predicting our experimental results.

To resolve this shortcoming, we reconsidered the heterodimer nonlinear optical response, emphasizing the extreme subwavelength scale of the system at hand. This reconsideration led us to propose a model based on the quasistatic near-field interaction between oscillating bodies of charge at each particle composing the dimer. According to this view, the optical nonlinearity does not stem from each particle on its own, which by itself is almost nonexistent due to inversion symmetry. Alternatively, we assume nonlinearity emerges from the Coulomb-like force between the oscillating charges at the different parts of the dimer, which dominates the inter-particle interaction for a sufficiently small structure. Our model successfully reproduced the measured SHG, thus, raising the possibility of a yet unknown type of nonlinear optics.

**Results:**

Fabrication of nanostructures made from different metals is possible by the lithographic patterning of each metal, but is, however, extremely challenging due to the required layer alignment. We, on the contrary, overcame this adversity by combining nanoimprint lithography and double-angel evaporation of two metals. Here, the shadowing effect of the angle evaporation allows the production of metallic heterodimers within a single lithographic cycle. The entire fabrication process shows in Figure 1(a): We first nanoimprinted thermal resist on either a silicon or glass substrates with periodic features of $\sim 20\ nm$ in diameter, following by angle evaporation of titanium and resist over-etch through the holes formed in titanium hard-mask using oxygen plasma.[23] We then evaporated two metals at opposite angles and made a resist liftoff-process to obtain the dimer arrays. The unique elegance of this fabrication approach is the total array density, ranging 25 to 400 dimers per squared micron, controlled by design of the nanoimprint mold, while the evaporation angle determined the spacing between the nano-dots within each dimer. The choice of metal at each angular evaporation step determined the formation of a homodimer (Ag-Ag, Au-Au) or heterodimer (Ag-Au).

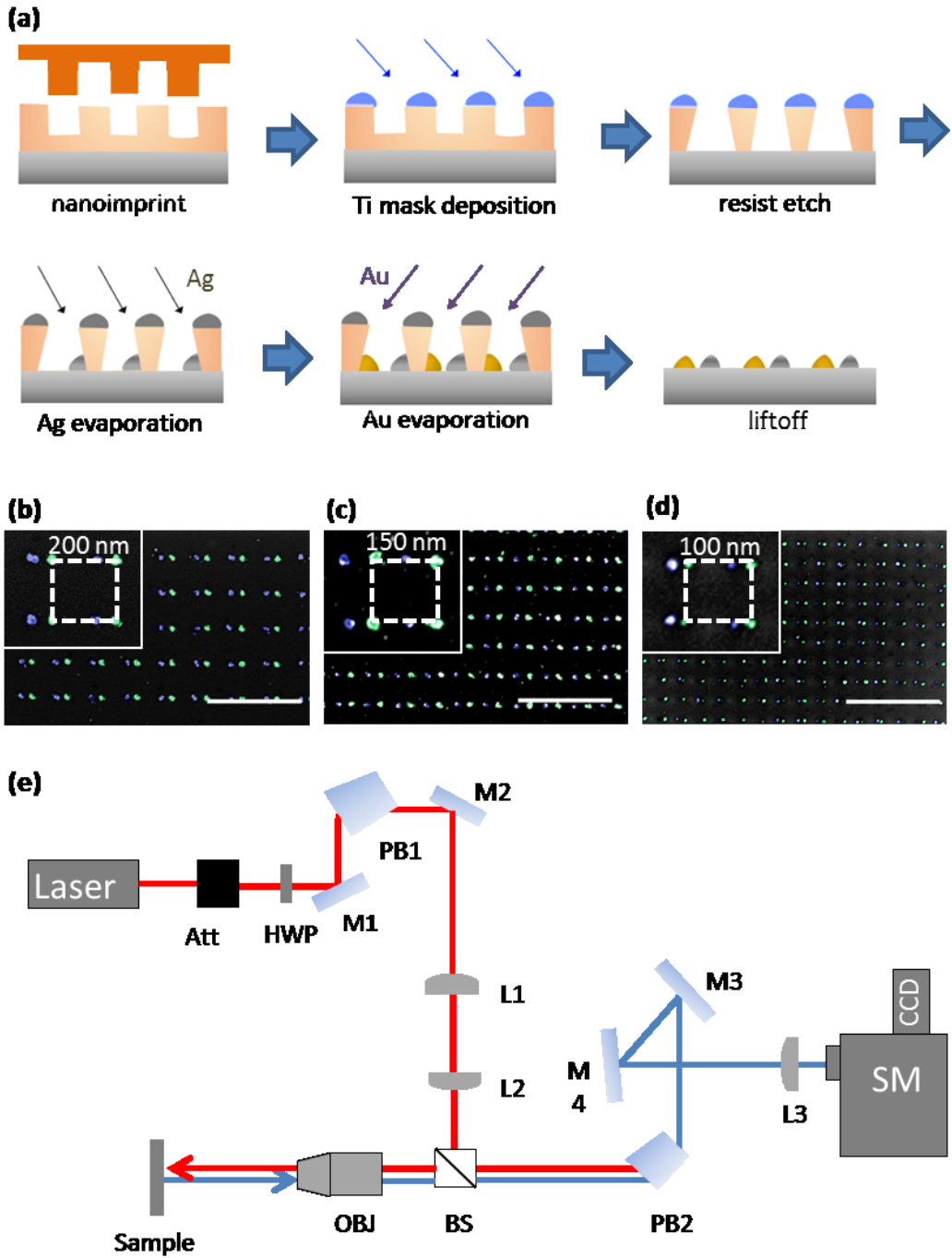

**Figure 1:** (a) Scheme of the fabrication process. (b) – (d) False-colored SEM images of nanofabricated arrays of silver-gold dimers, with the densities of 25, 45, and 100 dimers per squared micron, respectively. The scale bar at each image: 500 $nm$. Insets show a magnified unit-cell of the respective dimer arrays with the corresponding dimensions. (e) Schematic

depiction of the experimental setup showing the excitation laser, attenuator (Att), half-wave plate (HWP), mirrors (M1-M4), lenses (L1-L3), Pellin-Broca prisms (PB1, PB2), objective (OBJ), beam splitter (BS), and spectrometer (SM) with a CCD camera.

The SHG from Ag-Au heterodimers and Au-Au homodimers arrays at densities of 25, 45, 100, 200, and 400 dimers per squared micron were measured using the setup illustrated in Fig. 1(b): A tunable Ti:Sapphire laser with 150 $fs$ pulse duration at 80 $GHz$ repetition rate served as the source. An optical attenuator (Att) maintained time-averaged laser power at 2 $W$, which was well below the sample's damage threshold. Half-wave plate (HWP) was used to align the polarization along the dimers axis. A Pellin-Broca prism removed any residual laser luminescence at the designated SHG wavelength. Afterward, the expanded beam was focused onto the sample using a 40X objective (0.66 NA). SHG from the samples, along with the reflected laser light, were gathered by the selfsame objective lens to be directed down a second Pellin-Broca prism, this time aligned to reject the laser light off the beam-path. Finally, a spectroscopic cooled camera was used to register the resulting SHG for excitation wavelength ranging from 780 to 920 $nm$. The two PB prisms were realigned to perform their designated function for each nominal wavelength.

Figure 2(a) shows typical SHG spectra of the Ag-Au samples from an 800 $nm$ laser excitation linearly polarized along the dimer axis (the axis connecting the center of the two particles). The depicted spectral lines are normalized by the respective sample density to highlight the per unit heterodimer response. The 25 heterodimers per squared micron sample have the highest per-heterodimer SHG. On the contrary, it is the dens-most samples that gave the least per-heterodimer SHG (not shown), which is due to the higher level of symmetry the denser sample possesses, where inter-dimer and intra-dimer spacing become almost identical. We, therefore, choose to work with the 100 dimers per squared micron samples, which displayed the right balance between total SHG and per dimer SHG.

SHG spectra from the 100 units per squared micron homodimer and heterodimer samples were captured for different excitation wavelength but always with linear polarization oriented along the dimer axis. Figure 2(b) shows the average SHG as a function of the wavelength for the heterodimer and homodimers under consideration; error bars depict standard deviation of no less than three different sample locations. The heterodimer has significantly larger SHG compared to the two homodimer samples. Also, the heterodimer displays a distinctive peak that is absent from the homodimers' response.

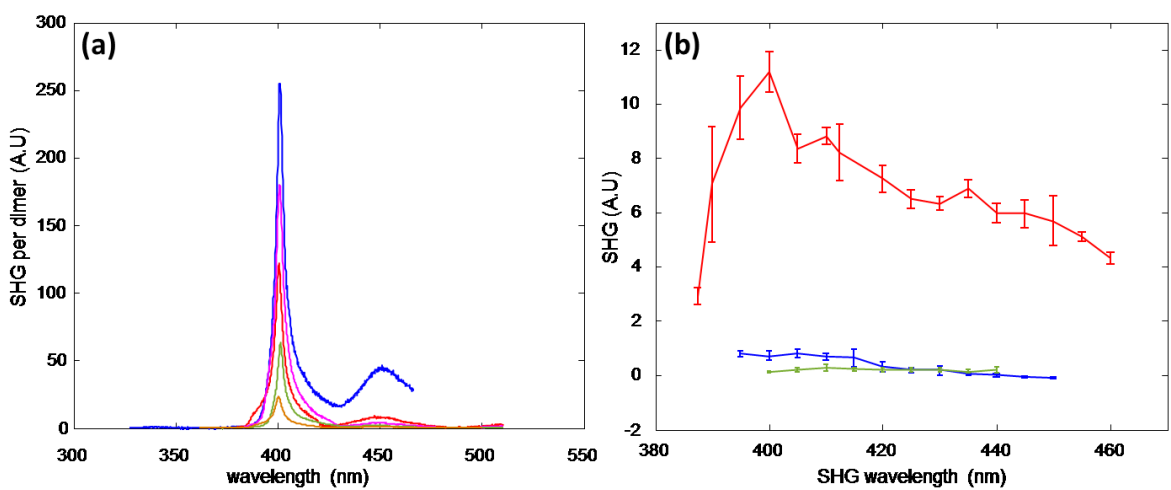

**Figure 2**: (a) Per dimer SHG from the different samples for $800\ nm$ laser excitation. blue: $25\ \mu m^{-2}$; red: $45 \mu m^{-2}$; pink: $100\ \mu m^{-2}$; green: $20\ \mu m^{-2}$; orange: $400\ \mu m^{-2}$. (b) Spatially averaged SHG as a function of wavelength for the $100\ \mu m^{-2}$ samples. red: Ag-Au, blue: Ag-Ag; green: Au-Au.

To better understand these results, we simulated the far-field scattered SHG of homodimers and heterodimers using a commercial simulation tool. The dimers consisted of 40 $nm$ particles with a 10 $nm$ gap - see the 'Methods' section for more details. Figure 3(a) shows the simulated SHG scattering cross-section of Au-Ag heterodimers and Ag-Ag homodimers. One can see that, according to simulation, the SHG peaks when its wavelength coincides with a resonance of the *linear* dimer response. Specifically, the ~360 $nm$ SHG peak emerges due to the excitation of the interband transitions of the Ag particle at this wavelength, well understood within the *conventional-view* of such processes.[18,22]. It is understood, therefore, that the Ag-Ag homodimer is expected to produce the highest level of SHG since it lacks one Au particle that inflicts more losses. Most importantly, the simulated SHG does not show the heterodimer measured spectral peak. Figure 3(b) compares the measured and simulated results for the Ag-Ag homodimer. The two can be considered to be in good agreement. Simulating a $2 \times 2$ super-lattice of dimers at selected wavelengths produced similar results, thus negating the role of inter-dimer interaction as a possible source for the observed spectra. Most importantly, however, as far as the Ag-Au heterodimer is concerned, we conclude that its nonlinear optical response is beyond the grasp of what can be considered the conventional-view of such processes.

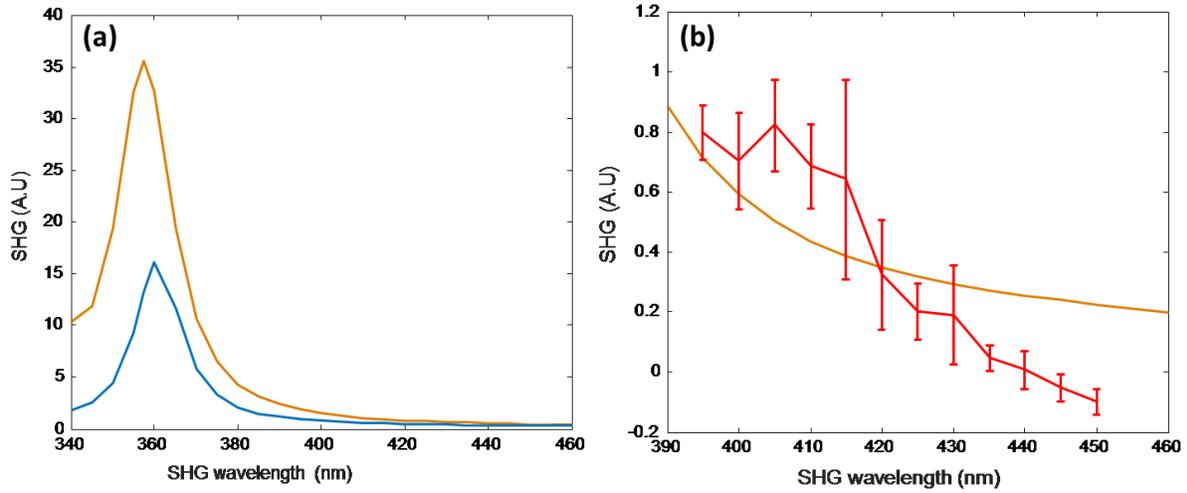

**Figure 3:** (a) Full-wave simulation for the Au-Ag heterodimer and the Ag-Ag homodimer in blue and orange, respectively. (b) Comparison between the simulated (red) and measured (blue) SHG results for Ag-Ag homodimer.

To understand what causes the unique heterodimer second-harmonic response, we need to re-asses the consequences of dealing with a relatively complex system whose dimensions are in the deep subwavelength scale. After all, each particle of the heterodimer is ~30 $nm$ in size in our case, which is comparable to the electromagnetic penetration depth into the respective metals at the frequency range of interest. Therefore, the electromagnetic field in the bulk of the particle is uniform in magnitude and phase. As a result, charges oscillate uniformly throughout the volume of the particle, without appreciable spatial variations. The optical response of each particle, in this case, is well described by a harmonic oscillator.[24,25] The other significant heterodimer character is its small inter-particle spacing, which is ~36 $nm$ in our case. This small gap leaves no room for appreciable phase retardation at the free-space wavelength of interest, which is no less than 350 $nm$ in our case. The interaction between particles is, therefore, electrostatic. Thus, as a result of the deep subwavelength dimensions, charges oscillating along the heterodimer axis interact across its gap. What would be the nature of this interaction? Since it is electrostatic, and since charge oscillations in each of the particles are spatially uniform, we expect Coulomb-like form:

$$\frac{Q^2}{D(t)^2}$$

Here $Q$ is a fixed amount of charge with $D(t)$ the instantaneous separation from its neighbor. If this is indeed the case, then, the nonlinearity of our system arises from the fact that the *dynamic variable appears squared at the denominator of the interaction term*. After all, if $D(t)^{-2} \propto (1 \pm x(t))^{-2}$ then its power series expansion, $\sum_n n \, x(t)^n$, is nonlinear. This view is in sharp contrast with the usual treatment of such cases. There, Coulombic interaction but with fixed interparticle separation is considered such that a strictly linear response emerges.[26–28]

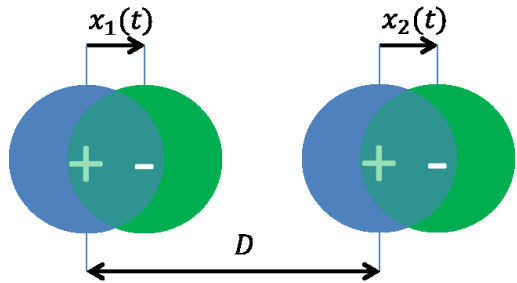

**Figure 4:** The system under consideration: Static positive charges are shown in blue, while negative mobile charges are in green. The parameter $D$ is the fixed separation between positive charges. The instantaneous shift of the negative charge relative to its respective static positive one give the dynamic variables $x_{1,2}$.

Let us consider the system in Fig. 4: The view mentioned above, namely of rigid charge oscillations and Coulomb-like interaction, is described by the following coupled dynamic equations:

$$\begin{cases} \ddot{x}_1 + 2\Gamma_1 \dot{x}_1 + \omega_1^2 x_1 = -\frac{e}{m_e} E_P(t) + \frac{k_e e^2 N_2}{m_e}\left[\frac{1}{(D-x_1)^2} - \frac{1}{(D-x_1+x_2)^2}\right] \\ \ddot{x}_2 + 2\Gamma_2 \dot{x}_2 + \omega_2^2 x_2 = -\frac{e}{m_e} E_P(t) - \frac{k_e e^2 N_1}{m_e}\left[\frac{1}{(D+x_2)^2} - \frac{1}{(D-x_1+x_2)^2}\right] \end{cases} \quad (1)$$

The instantaneous separation of negative charge in each particle relative to its static positive core is the dynamic variable $x_{1,2}$. The attraction between opposite charges within each particle gives the natural frequency $\omega_{1,2}$. Likewise, the damping rate of these natural oscillations is $\Gamma_{1,2}$. Any possible nano-scale optical confinement effect of such small particles is accounted for by these two phenomenological terms. Therefore, each particle is considered to have a linear response to the external force, where $E_P(t)$ is the electric field, $e$ is the fundamental charge, and $m_e$ is the electron mass.

There is, however, another force that acts on the oscillating negative charge of a given particle, which is the electrostatic forces from the neighboring particle. These forces are the Coulomb-like coupling terms enclosed in square brackets on the right-hand side of the equations. The first and second terms in each bracket denote, respectively, the interaction with the static positive or the oscillating negative charges across the heterodimer gap. Here, the number of charges in each particle is $N_1$ and $N_2$, the Coulomb constant is $k_e = \frac{1}{4\pi\varepsilon_0}$, and the vacuum permittivity in SI units is $\varepsilon_0$.

The proposed dynamic model has an intricate dependence on its parameters: $\omega_{1,2}$, $\Gamma_{1,2}$, $N_{1,2}$, $D$, and the magnitude of $E_p$. The exploration of which is, however, outside the scope of this article.

Alternatively, we focus on the leading order effects by noting that the charge oscillations $x_{1,2}$ are expected to be small compared to the inter heterodimer spacing $D$, which allows a series expansion of the coupling term up to second-order in powers of $\frac{x_{1,2}}{D}$. Assuming harmonic pump, $E_P = E_0 \sin(\omega t)$, and a trial solution of the form: $x_{1,2} = x_{1,2}^{(0\omega)} + x_{1,2}^{(1\omega)} \sin(\omega t + \varphi_{1,2}^{(1\omega)}) + x_{1,2}^{(2\omega)} \sin(2\omega t + \varphi_{1,2}^{(2\omega)})$, one receives, upon isolation of distinct harmonics and neglecting small-contributing terms, the following expressions for the magnitude and phase of the fundamental harmonics:

$$\begin{cases} x_{1,2}^{(1\omega)} = -\frac{e}{m_e} \frac{E_0}{\sqrt{(\omega_{1,2}^2 - \omega^2)^2 + 4\Gamma_{1,2}^2 \omega^2}} \\ \tan(\varphi_{1,2}^{(1\omega)}) = -\frac{4\Gamma_{1,2}\omega}{\omega_{1,2}^2 - \omega^2} \end{cases}, \quad (2)$$

Equation (2) reminisces the hybridization effect. [26–28] For simplicity, however, and since this is not the focus of this article, the impact of $x_2^{(1\omega)}$ on $x_1^{(1\omega)}$, and vise versa, have been neglected. Likewise, for the second-harmonic oscillations, we obtain:

$$\begin{cases} x_{1,2}^{(2\omega)} = \frac{3e^2}{m_e D^3} \sqrt{\frac{\left(N_{2,1} x_{1,2}^{(1\omega)} x_{2,1}^{(1\omega)}\right)^2 + \left(\frac{2}{9} N_{2,1} x_{2,1}^{(1\omega)^2}\right)^2 - N_{2,1}^2 x_{1,2}^{(1\omega)} x_{2,1}^{(1\omega)^3} \cos(\varphi_{1,2}^{(1\omega)} - \varphi_{2,1}^{(1\omega)})}{(\omega_{1,2}^2 - 4\omega^2)^2 + 16\Gamma_{1,2}^2 \omega^2}} \\ \tan\left(\frac{\varphi_{1,2}^{(2\omega)}}{2}\right) = -\frac{(A_{1,2} - B_{1,2}) + \sqrt{2A_{1,2}^2 + 2B_{1,2}^2 - C_{1,2}^2}}{(A_{1,2} + B_{1,2} + C_{1,2})} \end{cases}, \quad (3)$$

where:

$A_{1,2} = \frac{e^2 N_{2,1}}{D^5 m_e} \left[ 3 x_{1,2}^{(1\omega)} x_{2,1}^{(1\omega)} \cos(\varphi_{1,2}^{(1\omega)} + \varphi_{2,1}^{(1\omega)}) - 1.5 x_{2,1}^{(1\omega)^2} \cos(2\varphi_{2,1}^{(1\omega)}) \right]$;

$B_{1,2} = \frac{e^2 N_{2,1}}{D^5 m_e} \left[ 3 x_{1,2}^{(1\omega)} x_{2,1}^{(1\omega)} \sin(\varphi_{1,2}^{(1\omega)} + \varphi_{2,1}^{(1\omega)}) - 1.5 x_2^{(1\omega)^2} \sin(2\varphi_1^1) \right]$;

$$C_{1,2} = \frac{1}{D^2}\left[\left(\omega_{1,2}^2 - 4\omega^2\right) - 4\Gamma_{1,2}\omega\right]x_{1,2}^{(2\omega)}.$$

The above expressions are an approximate analytic solution, which is valid at relatively low external deriving force. To be able to compare Eq. (2) and (3) to our measured data, we assume that far-field scattered SHG is proportional to the square of the second-harmonic radiating dipole, namely: $I_{2\omega} \propto [N_1 x_1^{2\omega} \cos(\varphi_1^{2\omega}) + N_2 x_2^{2\omega} \cos(\varphi_2^{2\omega})]^2$, where $I_{2\omega}$ is the on-axis far-field intensity. For the homodimer case, where $\omega_1 = \omega_2$ and $\Gamma_1 = \Gamma_2$, Eq. (2) and (3) gives $x_1^{2\omega} = x_2^{2\omega}$ and $\varphi_2^{2\omega} = \varphi_1^{2\omega} + \pi$ such that $I_{2\omega} = 0$ emerges, as required from symmetry considerations, which expresses the so-called silencing effect in our model.[21,22,29]

Figure 5 shows a fit of our model to the measured results. Apart from a trivial scaling, the fitting parameters that emerged in this case were: $\omega_1 = 3.715 \times 10^{15}\ s^{-1}$ corresponding a $507\ nm$ Au particle resonance, and $\omega_2 = 4.918 \times 10^{15}\ s^{-1}$ corresponding a $383\ nm$ Ag particle resonance. Also, $\Gamma_1 = 3.098 \times 10^{14}\ s^{-1}$ and $\Gamma_2 = 1.757 \times 10^{14}\ s^{-1}$ have emerged for the damping in the Au and Ag particles, respectively. The good agreement between the proposed model and measured data suggests the existence of nonlinear interaction of the prescribed manner. The fact that this agreement was obtained for a realistic set of fitting parameters, considering the system at hand, re-enforces this impression.

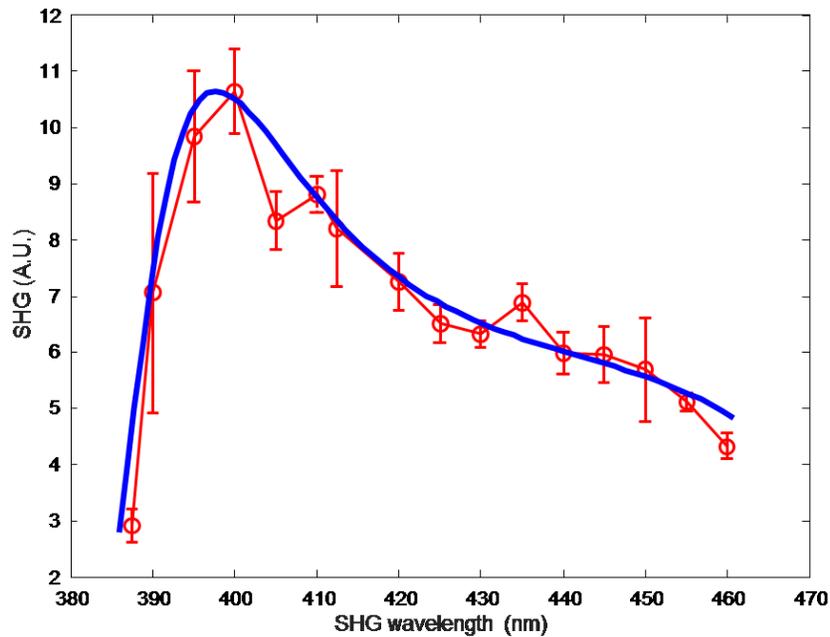

**Figure 5:** Calculated (blue) and measured (red) peak SHG of an Ag-Au heterodimer.

**Summary and conclusions:**

The SHG from the deep subwavelength sized Ag-Au heterodimer has shown to be markedly different than the expectation from the conventional theoretical view in that case. On the contrary, the SHG from Ag-Ag homodimer was well-reproduced by simulating this standard view. This fact has led us to conclude that the heterodimer somehow possesses a yet unknown source of nonlinear optical activity. Re-evaluating the situation led us to adopt a Coulomb-like interaction term with fixed amounts of charges and dynamic separation between them. We, therefore, find that the optical nonlinearity of the heterodimer stems not from the particles themselves but form the unique nature of the interaction between them. The nonlinearity that emerges, in this case, encompasses all harmonics and possibly even beyond. Aiming only to reproduce the measured SHG at this stage, we derive an approximate expression for the first and second harmonics. These expressions successfully reproduce the measured reality with realistic values for the fitting parameters.

Based on the success of the abovementioned fit, we now aim to estimate what could be the maximal SHG from this particular heterodimer system. A brief inspection of Eq. (3) shows that the second-order oscillations have their poles at the fundamental and second-harmonic frequencies. Let us, therefore, fix the natural frequency and the dumping rate of one oscillator to be that of the fitted Au particle. Also, let us consider the excitation laser frequency to be the natural frequency of the Au particle. Figure 6 then shows with orange-line the calculated relative SHG that emerges from scanning the frequency of the remaining particle while keeping the fitted value of the Ag particle dumping rate. The results are then normalized by the intensity that emerged once fitting the experimental data, as shown in Fig. 5. Peak SHG emerges when the second particle resonates at twice the natural frequency of the first particle, which is also the excitation frequency in this case. This maximum is almost 150 times more intense than what was measured experimentally, which was at a frequency ratio of 1.32 between the Au and Ag particles, as indicated by the orange arrow. There, the graph assumes a value of one. So far, an inter-particle spacing of 30 $nm$ was considered. Equation (3), however, indicates that the SHG is $\propto D^{-3}$. Therefore, the blue line in the figure shows that if interparticle spacing would somehow reduce by only 10 $nm$, SHG would become almost 4000 times more intense than what was measured, if natural frequencies are at an optimal ratio of 1: 2 and the excitation is at the lower one. The potential of the mechanism mentioned above to act as an immense nonlinear optical source is, therefore, shown.

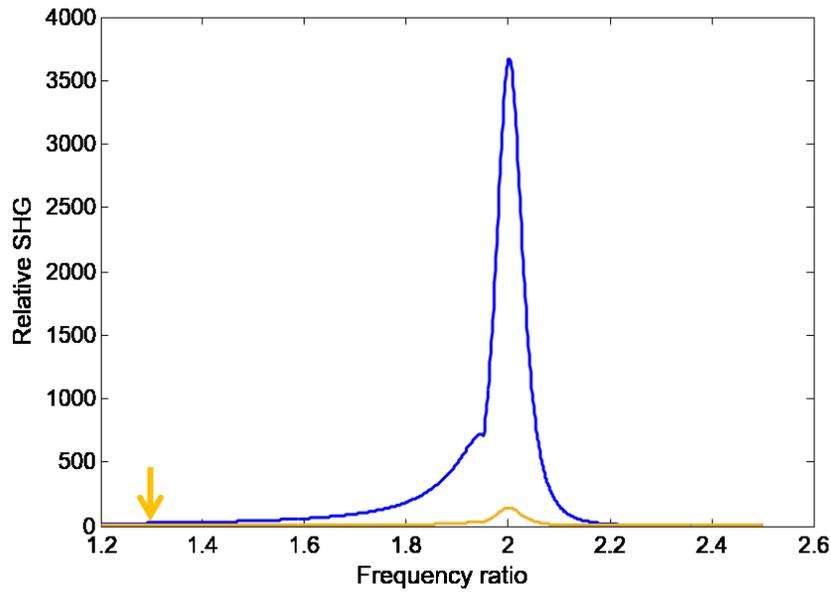

**Figure 6:** SHG as a function of the relative frequency of the two oscillators plotted relative to the maximum fitted SHG of Fig. 5. Orange-line shows what emerges from maintaining **30 $nm$** inter-dimer spacing while the blue line shows what could be if this spacing would be **20 $nm$** instead. The experimental frequency ratio is at a value of **1.32**, which is indicated by the orange arrow.

The fact that such a simple model can fatefully reproduce the measured reality is an impressive feat. It cannot, however, replace a simulation of the given scenario that considers Maxwell's equation at their full and the emergent charge-dynamics. In this regard, impressive progress has been achieved lately with embedding nonlocal effects in the simulated optical response of nanoscaled systems.[30–32] These, however, considers only the free-charges of metals, usually within the hydrodynamic model. On the contrary, the metals optical response at the near infrared and visible spectrum is influenced and sometimes dominated by bound charges in the form of inter- and intra-band transitions.[9,33,34] Unfortunately, inter and intraband transition are not accounted for by the above mentioned contemporary simulation tools. Our model holds a particular strength in this respect since it considers the linear optical response of each metal particle in a purely phenomenological manner. Therefore, inter and intraband transition, hydrodynamic, and hybridization effects are all equally accounted for by the phenomenological natural-frequencies and damping rates of our model. The possible role of bound charges in the reported results raises the possibility of similar findings but for heterodimers made from semiconducting quantum-dots instead of metals. Lastly, it is not clear at this stage what is the strength of this nonlinear interaction-based source, especially compared to other nonlinear sources such as natural crystals and plasmonic meta-materiels. Other open questions concern the role of the mechanism mentioned above in different kinds of second-order nonlinear processes, such as sum and difference frequency generation, optical-rectification, as well as higher-order nonlinear optical processes.

**Corresponding Author**

*aviniv@bgu.ac.il

**Author Contributions**

The manuscript was written through the contributions of all authors. All authors have approved the final version of the document. ‡These authors contributed equally.

**Funding Sources**

A.N is grateful for the partial support of the Israel Science Foundation (ISF) no. 152/11 and the Adelis Foundation. Y. S. was partially supported by Israel Science Foundation (ISF) no. 899/16.

**Methods**

**Fabrication:** We fabricated the dimer arrays by nanoimprint lithography. For that purpose, we first produced a nanoimprint mold using electron beam lithography of hydrogen silsesquioxane (HSQ, XR-1541, Dow Corning), which was diluted in Methyl Isobutyl Ketone (MIBK) and spin-coated on Silicon substrate to produce 20 nm thick film. We patterned HSQ by exposing it in Raith e-LINE e-beam tool at the acceleration voltage of 20 kV, developing in 0.26 N TMAH solution (AZ726, Rohm and Hass) for 2 min, rinsing with DI water, and drying with nitrogen. We then annealed the mold at 450 °C for 1 hour in a nitrogen atmosphere and coated it with an anti-adhesive agent (NXT-110, Nanonex NXT-110). We used the mold to imprint dimer arrays on glass coverslips or Silicon substrates, using spin-coated PMMA (50 K, Microchem) with a thickness of 40 nm as a resist. For the imprint, we used a commercial nanoimprint tool (NX-B100, Nanonex), with the following parameters: $450\ psi$, $180°\ C$, and imprinting time of 4 minutes. We then evaporated $10\ nm$ thick Ti mask onto the imprinted PMMA film, while tilting the substrate by 30° to ensure that the mask covers only the top PMMA surface but not the bottom of the imprinted features. To remove the excess of the resist and produce a wide undercut in the Ti mask, we etched PMMA through the formed Ti mask with oxygen plasma (Corial 200 IL, 10 mTorr, and 10 sccm oxygen flow of). For heterodimer arrays, we sequentially evaporated Au ($7\ nm$) and Ag ($7\ nm$) though the formed mask while tilting the substrate at opposite angles of 60° and -60°, respectively. For homodimer arrays, we used accordingly used the same metal for both evaporations. To enhance the adhesion of Au and Ag to Silicon or glass, we evaporated a few nm of Ti before the deposition of each metal. Finally, we performed a liftoff of PMMA in boiling acetone. We covered the fabricated samples with a thick layer of PMMA by spin-coating and baking to prevent oxidation of the Ag.

**Characterization:** All samples have been studied using a tunable laser source (Short pulse tunable Ti:Sapphire laser, 150 fs pulse at 80 GHz repetition rate; Model: Chameleon Ultra II; Manufacturer: Coherent), generating an SHG signal that was recorded by a spectrometer (Andor Shamrock 303i with iDus 420 UV enhanced CCD camera).

**Simulations:** A single heterodimer, consisting of a 40 nm Au and Ag particles with 10 *nm* interparticle spacing, was simulated using the known material properties.[35] The polarization of the excitation light was considered to be parallel to the dimer axis. The simulation utilized scattering boundary conditions and perfectly matched layers. Bulk nonlinearity was mapped to the surface such that all non-linearity was assumed to arise at the surface.[36] The second-harmonic polarization at the surface of the particles was:

$$P_{s,\perp}^{(2\omega)} = \varepsilon_0 \chi_{s,\perp,\perp,\perp}^{(2)} E_\perp^{(1\omega)}(r_\parallel) E_\perp^{(1\omega)}(r_\parallel) \hat{r}_\perp,$$

where $E_\perp^\omega$ is the normal electric field at the inner side of the metal-dielectric interface. The second-order susceptibility is given by the relation:[15,20]

$$\chi_{s,\perp,\perp,\perp}^{(2\omega)} = \frac{1}{4}[\varepsilon_r(\omega) - 1]\frac{e\varepsilon_0}{m_e \omega^2}.$$

Note that additional elements of the nonlinearity tensor are neglected. The polarization from above leads to the following surface current just outside the metal:

$$J_{s,\perp}^{(2\omega)} = -2i\omega P_{s,\perp}^{(2\omega)}.$$

The non-uniform normal component of the surface current creates a discontinuity in the tangential electric field ($E_\parallel$) at the metal interface implemented as an (artificial) surface magnetic current density:[37]

$$J_{m,S,\parallel}^{(2\omega)} = \frac{1}{\varepsilon}\hat{r}_\perp \times (\nabla_\parallel P_{S,\perp}^{(2\omega)}),$$

where $\varepsilon$ is the background permittivity taken here as a vacuum.

Boundary Conditions for Second-Harmonic Generation at Metal-Dielectric Interfaces. *ArXiv* **2017**, 1–9.